\documentclass[a4paper,10pt,eqno]{article}
\usepackage{theorem}
\usepackage{latexsym,amssymb,amsfonts,amsmath}
\usepackage{cite}
\usepackage{color}
\newtheorem{thm}{Theorem}[section]
\newtheorem{lem}[thm]{Lemma}
\newtheorem{cor}[thm]{Corollary}

\newtheorem{rem}[thm]{Remark}
\newtheorem{ass}[thm]{Assumption}

\theoremheaderfont{\scshape}

\newcommand{\CM}{\mathbb{C}}

\newcommand{\PM}{\mathbb{P}}

\newcommand{\qed}{\hfill $\Box$}

\newcommand{\ket}[1]{|#1\rangle}
\newcommand{\bra}[1]{\langle#1|}

\title{{\Large {\bf Periodicity for space-inhomogeneous quantum walks on the cycle}}}
\author{
{\small Toshiyuki Arai}\\
{\scriptsize Department of Applied Mathematics, 
Faculty of Engineering, 
Yokohama National University}\\
{\scriptsize Hodogaya, Yokohama 240-8501, Japan}\\
{\scriptsize e-mail: t04tttt@gmail.com }\\
\\
{\small Choon-Lin Ho}\\
{\scriptsize Department of Physics, Faculty of Core Research,
Ochanomizu University}\\
{\scriptsize Bunkyo-ku, Tokyo 112-8610 , Japan}\\
{\scriptsize Department of Physics, 
Tamkang University \footnote{Permanent address}}\\
{\scriptsize Tamsui 251, Taiwan (R.O.C.)}\\
{\scriptsize e-mail: hcl@mail.tku.edu.tw}\\
\\
{\small Yusuke Ide
\footnote{To whom correspondence should be addressed. E-mail: ide@kanagawa-u.ac.jp}}\\
{\scriptsize Department of Information Systems Creation, 
Faculty of Engineering, 
Kanagawa University}\\
{\scriptsize Kanagawa, Yokohama 221-8686, Japan}\\
{\scriptsize e-mail: ide@kanagawa-u.ac.jp}\\
\\
{\small Norio Konno}\\
{\scriptsize Department of Applied Mathematics, 
Faculty of Engineering, 
Yokohama National University}\\
{\scriptsize Hodogaya, Yokohama 240-8501, Japan}\\
{\scriptsize e-mail: konno@ynu.ac.jp}\\
}
\date{\empty }
\begin{document}
\maketitle

\par\noindent
\begin{small}
\par\noindent
{\bf Abstract}
\newline 
In this paper, we consider periodicity for space-inhomogeneous quantum walks on the cycle. For isospectral coin cases, we propose a spectral analysis. Based on the analysis, we extend the result for periodicity for Hadamard walk to some isospectral coin cases. For non-isospectral coin cases, we consider the the system that uses only one general coin at the origin and the identity coin at the other sites. In this case, we show that the periodicity of the general coin at the origin determines the periodicity for the whole system.
\end{small}

\setcounter{equation}{0}
\section{Introduction}
In the last two decades, the theory of quantum walk (QW) has bees extensively studied by many researchers. There exist good reviews for this development, for example, Kempe \cite{Kempe2003}, Kendon \cite{Kendon2007}, Venegas-Andraca \cite{VAndraca2008, VAndraca2012}, Konno \cite{Konno2008b}, Manouchehri and Wang \cite{ManouchehriWang2013},  and Portugal \cite{Portugal2013}. In the present paper, we focus on periodicity of the time evolution operator of two-state discrete-time QWs (DTQWs) on the cycle graph. The periodicity of the Hadamard walk on the cycle graph was determined by Dukes \cite{Dukes2014} and Konno et al. \cite{KonnoShimizuTakei2015}. Note that the word periodicity is also used in the theory of perfect state transfer \cite{Godsil2012, Coutinho2014} but we consider little bit stronger version of periodicity in this paper.

The rest of this paper is organized as follows. In Sect.\ 2, we give the definitions of our DTQWs and periodicity. Sections 3 and 4 are devoted to spectral analysis of the time evolution operator of our DTQWs. We note that the spectral analysis is viewed as a generalization of that of Segawa \cite{Segawa2013}. Corollary \ref{cor:UMSHadamard} is an extension of the results given by Dukes \cite{Dukes2014} and Konno et al. \cite{KonnoShimizuTakei2015} for space-inhomogeneous coin cases.  In Sect.\ 5, we deal with periodic arranged coin cases which is motivated by Chou and Ho \cite{ChouHo2014}.

\section{Definition of the DTQWs on the cycle graph}\label{sect:def}
In this paper, we consider DTQWs on the cycle graph $C_{n}=(V_{n}, E_{n})$ with the vertex set $V_{n}= \{0,1,\ldots ,n-1\}$ and the edge set $E_{n}= \{(i,i+1):i\in V_{n}\ (\!\!\! \mod n)\}$. The Hilbert space of DTQWs is defined by $\mathcal{H}_{n}= \mathrm{span}\{\ket{i,L},\ket{i,R}:i\in V_{n}\}$ with state vectors $\ket{i,J}=\ket{i}\otimes \ket{J}\ (i\in V_{n}, J\in \{L,R\})$ given by the tensor product of elements of two orthonormal bases:  $\{\ket{i}:i\in V_{n}\}$ for position of the walker, and $\{\ket{L}={}^T [1,0], \ket{R}={}^T [0,1]\}$ for the chirality (direction) of the motion of the walker. Here ${}^T \!\!A$ denotes the transpose of a matrix $A$. 

Now we define two types of time evolution operators $U^{MS}_{n}=S_{n}^{MS}\mathcal{C}_{n}$ and $U^{FF}_{n}=S_{n}^{FF}\mathcal{C}_{n}$ on $\mathcal{H}_{n}$ with the coin operator $\mathcal{C}_{n}$, the moving shift operator $S^{MS}_{n}$ and the flip-flop shift operator $S^{FF}_{n}$ defined as follows:
\begin{align*}
\mathcal{C}_{n}&=\sum_{i=0}^{n-1}\ket{i}\bra{i}\otimes C_{i},\\
S_{n}^{MS}\ket{i,J}&=
\begin{cases}
\ket{i+1,R}\ (\!\!\! \mod n)&\text{if}\ \ J=R,\\ 
\ket{i-1,L}\ (\!\!\! \mod n)&\text{if}\ \ J=L,
\end{cases}\\
S_{n}^{FF}\ket{i,J}&=
\begin{cases}
\ket{i+1,L}\ (\!\!\! \mod n)&\text{if}\ \ J=R,\\ 
\ket{i-1,R}\ (\!\!\! \mod n)&\text{if}\ \ J=L,
\end{cases}
\end{align*}
where $C_{i}\ (i=0,\ldots ,n-1)$ are $2\times 2$ unitary matrices. 

Let $X_{t}^{(n)}\in V_{n}$ be the position of our quantum walker driven by the time evolution operator $U_{n}$ ($=U_{n}^{MS}$ or $U_{n}^{FF}$) at time $t$. The probability that the walker with an initial state (unit vector) $\ket{\psi}\in \mathcal{H}_{n}$ is found at time $t$ and the position $x$ is defined by 
\begin{eqnarray*}
\PM_{\ket{\psi}}(X_{t}^{(n)}=x)=\left\lVert \left(\bra{x}\otimes I_{2}\right)U^{t}\ket{\psi}\right\rVert^{2}.
\end{eqnarray*}
In this paper, we consider periodicity of the DTQWs. In order to define periodicity, we use the following notation:
\begin{align}
T_{n}(U)
=\inf \left\{t:U^{t}=I_{n}\otimes I_{2}\right\}.\label{def:T_{n}(U)}
\end{align}
We will investigate the period $T_{n}(U_{n}^{MS})$ and $T_{n}(U_{n}^{FF})$. We should remark the following fact:
\begin{rem}\label{rem:spec}
Let $\lambda _{1},\ldots ,\lambda _{2n}$ be the eigenvalues of the time evolution operator $U_{n}$ ($=U_{n}^{MS}$ or $U_{n}^{FF}$) then $U_{n}^{t}=I_{n}\otimes I_{2} \iff \lambda _{1}^{t}=\cdots =\lambda _{2n}^{t}=1$.
\end{rem}
By Remark \ref{rem:spec}, the spectral structure of the time evolution operators are important. Here we show a connection between $\mathrm{Spec}\ U_{n}^{MS}$ and $\mathrm{Spec}\ U_{n}^{FF}$.
\begin{lem}\label{lem:spec}
Let $\sigma _{x}=\ket{R}\bra{L}+\ket{R}\bra{L}$ and $\mathcal{C}_{n}\sigma_{x}=\sum_{i=0}^{n-1}\ket{i}\bra{i}\otimes C_{i}\sigma_{x}$. We denote $U_{n}^{MS}(\mathcal{C}_{n})=S^{MS}\mathcal{C}_{n}$ and $U_{n}^{FF}(\mathcal{C}_{n})=S^{FF}\mathcal{C}_{n}$. Then we have $\mathrm{Spec}\ U_{n}^{FF}(\mathcal{C}_{n})=\mathrm{Spec}\ U_{n}^{MS}(\mathcal{C}_{n}\sigma_{x})$
\end{lem}
{\bf Proof of Lemma \ref{lem:spec}.}

By the definition, we have $S_{n}^{FF}=(I_{n}\otimes \sigma_{x})S_{n}^{MS}$. Then by using $(I_{n}\otimes \sigma_{x})^{2}=(I_{n}\otimes I_{2})$, we obtain
\begin{align*}
U_{n}^{FF}(\mathcal{C}_{n})
&=
S^{FF}\mathcal{C}_{n}
=
(I_{n}\otimes \sigma_{x})S_{n}^{MS}\mathcal{C}_{n}
=
(I_{n}\otimes \sigma_{x})S_{n}^{MS}\mathcal{C}_{n}(I_{n}\otimes \sigma_{x})^{2}
=
(I_{n}\otimes \sigma_{x})S_{n}^{MS}\mathcal{C}_{n}\sigma_{x}(I_{n}\otimes \sigma_{x})\\
&=
(I_{n}\otimes \sigma_{x})U_{n}^{MS}(\mathcal{C}_{n}\sigma_{x})(I_{n}\otimes \sigma_{x}).
\end{align*}
This completes the proof.
\qed
 
Lemma \ref{lem:spec} shows that $T_{n}(U_{n}^{MS})=T_{n}(U_{n}^{FF})$ whenever we consider a pair of DTQWs defined by $U_{n}^{MS}(\mathcal{C}_{n}\sigma_{x})$ and $U_{n}^{FF}(\mathcal{C}_{n})$. Note that the coin operator $\mathcal{C}_{n}\sigma_{x}$ is given by exchanging column of all $C_{i}$ in $\mathcal{C}_{n}=\sum_{i=0}^{n-1}\ket{i}\bra{i}\otimes C_{i}$.

\section{Jacobi matrix}
Before we investigate periodicity of quantum walks defined in Sect.\ref{sect:def}, it is helpful to consider a related Jacobi matrix. Let $\nu _{1,i}, \nu _{2,i}$ and $\ket{w_{1,i}}, \ket{w_{2,i}}$ be the eigenvalues and the corresponding orthonormal eigenvectors of $C_{i}\ (i=0,\ldots ,n-1)$. We consider the spectral decomposition of each unitary matrix $C_{i}$ as follows:
\begin{align}
C_{i}
&=\nu _{1,i}\ket{w_{1,i}}\bra{w_{1,i}}+\nu _{2,i}\ket{w_{2,i}}\bra{w_{2,i}}\notag\\
&=\nu _{1,i}\ket{w_{1,i}}\bra{w_{1,i}}+\nu _{2,i}\left(I_{2}-\ket{w_{1,i}}\bra{w_{1,i}}\right)\notag \\
&=\left(\nu _{1,i}-\nu _{2,i}\right)\ket{w_{1,i}}\bra{w_{1,i}}+\nu _{2,i}I_{2},\label{specC}
\end{align}
where $I_{k}$ is the $k\times k$ identity matrix. Here we use the relation $I_{2}=\ket{w_{1,i}}\bra{w_{1,i}}+\ket{w_{2,i}}\bra{w_{2,i}}$ coming from unitarity of $C_{i}$. This shows that we can represent $C_{i}$ without $\ket{w_{2,i}}$. 

We define the $n\times n$ Jacobi matrix $J_{n}^{QW}$ for the DTQW as follows:
\begin{align}\label{defJacobiQW}
(J_{n}^{QW})_{i,j}=\overline{(J_{n}^{QW})_{j,i}}
=
\begin{cases}
\overline{w_{i}(R)}w_{j}(L) & \text{if $j=i+1\ (\! \! \mod n)$,}\\
0 & \text{otherwise,}
\end{cases}
\end{align}
where $\ket{w_{1,i}}={}^T [w_{i}(L),w_{i}(R)]$ and $\overline{z}$ means the complex conjugate of $z\in \CM$.  In this setting, the corresponding Jacobi matrix is the following:
\begin{align}\label{matJacobiQW}
J_{n}^{QW}=
\begin{bmatrix}
0 & \overline{w_{0}(R)}w_{1}(L) & & & \overline{w_{0}(L)}w_{n-1}(R)\\
\overline{w_{1}(L)}w_{0}(R) & 0 & \ddots & & \mbox{\smash{\huge\textit{O}}} & & \\
 & \ddots & \ddots & \ \ddots & \\
 & & \ddots & 0 & \overline{w_{n-2}(R)}w_{n-1}(L)\\
\overline{w_{n-1}(R)}w_{0}(L) & \mbox{\smash{\huge\textit{O}}} & & \overline{w_{n-1}(L)}w_{n-2}(R) & 0
\end{bmatrix}.
\end{align}
As we will point out at the next line of Eq.\ \eqref{eq:Uab}, each eigenvalue of $J_{n}^{QW}$ becomes inner product of two unit vectors. It means that $\textrm{Spec}(J_{n}^{QW})\subseteq [-1,1]$.
By direct calculation, we obtain the following lemma for the characteristic polynomial of the Jacobi matrix $J_{n}^{QW}$:
\begin{lem}\label{lem:polyJacobi}
Let 
\begin{align*}
K_{i,j}^{QW}(\lambda)=
\begin{bmatrix}
\lambda & -\overline{w_{i}(R)}w_{i+1}(L) & & & & & \\
-\overline{w_{i+1}(L)}w_{i}(R) & \lambda & \ddots & & \mbox{\smash{\huge\textit{O}}} & \\
 & \ddots & \ddots & \ddots & \\
 & & \ddots & \lambda & -\overline{w_{j}(R)}w_{j+1}(L)\\
\mbox{\smash{\huge\textit{O}}} & & & -\overline{w_{j+1}(L)}w_{j}(R) & \lambda
\end{bmatrix}.
\end{align*}
Then 
\begin{align}\label{eq:polyJacobi}
\det(\lambda I_{n}-J_{n}^{QW})
&=
\lambda \det(K_{1,n-2}^{QW}(\lambda ))\\ \notag
&-
|w_{0}(R)|^{2}|w_{1}(L)|^{2}\det(K_{2,n-2}^{QW}(\lambda ))-|w_{n-1}(R)|^{2}|w_{0}(L)|^{2}\det(K_{1,n-3}^{QW}(\lambda ))\\ \notag
&+
(-1)^{n}\cdot 2\Re \left(\prod_{i=0}^{n-1}\overline{w_{i}(R)}w_{i}(L)\right),
\end{align}
where $\Re (z)$ denotes the real part of $z\in \mathbb{C}$.
\end{lem}

In addition, we have 
\begin{align*}
 &\det(K_{i,j}^{QW}(\lambda))=\lambda \det(K_{i,j-1}^{QW}(\lambda))-|w_{j}(R)|^{2}|w_{j+1}(L)|^{2}\det(K_{i,j-2}^{QW}(\lambda)),\ (j\geq i+1),\\
 &\det(K_{i,i}^{QW}(\lambda))=\lambda ^{2} - |w_{i}(R)|^{2}|w_{i+1}(L)|^{2},\\
\end{align*}
with a convention $\det(K_{i,i-1}^{QW}(\lambda))=\lambda.$
This leads to the following lemma:
\begin{lem}\label{lem:Kreal}
$\det(K_{i,j}^{QW}(\lambda))$ is a polynomial with real coefficients. If we define $p_{i}=|w_{i}(R)|^{2}$ and $q_{i}=|w_{i}(L)|^{2}$ for $i\in V_{n}$ then the coefficients of $\det(K_{i,j}^{QW}(\lambda))$ are determined by $p_{i}, \ldots, p_{j}, q_{i}, \ldots ,q_{j+1}$.
\end{lem}

\section{Isospectral coin cases}

Now we give a framework of spectral analysis for DTQWs with flip-flop shift on $C_{n}$. In order to do so, we restrict the coin operator as follows:
\begin{ass}\label{ass:coinQW}
We assume all the local coins are isospectral. Thus we use
\begin{align}\label{SpecAnalcoinQWonPath}
\mathcal{C}_{n}
=
\sum_{i=0}^{n-1}\ket{i}\bra{i}\otimes \left\{(\nu _{1}-\nu _{2})\ket{w_{i}}\bra{w_{i}}+\nu _{2}I_{2}\right\}
,
\end{align}
as the coin operator.
\end{ass}

Let $\lambda _{m}\ (m=0,\ldots n-1)$ be the eigenvalues and $\ket{v_{m}}\ (m=0,\ldots n-1)$ be the corresponding (orthonormal) eigenvectors of $J_{n}^{QW}$. For each $\lambda _{m}$ and $\ket{v_{m}}$, we define two vectors
\begin{align*}
\mathbf{a}_{m}
&=
\sum_{i=0}^{n-1}v_{m}(i)\ket{i}\otimes \ket{w_{i}},\\
\mathbf{b}_{m}
&=
S_{n}^{FF}\mathbf{a}_{m},
\end{align*}
where $\ket{v_{m}}={}^{T}\left[v_{m}(0) \ldots v_{m}(n-1)\right]$. By using $(S_{n}^{FF})^{2}=I_{n}\otimes I_{2}$, it is easy to see that $\mathcal{C}_{n}\mathbf{a}_{m}=\nu_{1}\mathbf{a}_{m}$ and then $U_{n}^{FF}\mathbf{a}_{m}=\nu_{1}\mathbf{b}_{m}$. Also we have $\mathcal{C}_{n}\mathbf{b}_{m}=(\nu_{1}-\nu_{2})\lambda _{m}\mathbf{a}_{m}+\nu_{2}\mathbf{b}_{m}$ and $U_{n}^{FF}\mathbf{b}_{m}=\nu_{2}\mathbf{a}_{m}+(\nu_{1}-\nu_{2})\lambda _{m}\mathbf{b}_{m}$. So we have the following relationship:
\begin{eqnarray}\label{eq:Uab}
U_{n}^{FF}
\begin{bmatrix}
\mathbf{a}_{m}\\
\mathbf{b}_{m}
\end{bmatrix}
=
\begin{bmatrix}
0 & \nu_{1}\\
\nu_{2} & (\nu_{1}-\nu_{2})\lambda _{m}
\end{bmatrix}
\begin{bmatrix}
\mathbf{a}_{m}\\
\mathbf{b}_{m}
\end{bmatrix}.
\end{eqnarray}
We also obtain $|\mathbf{a}_{m}|=|\mathbf{b}_{m}|=1$ and the inner product $(\mathbf{a}_{m}, \mathbf{b}_{m})=\lambda _{m}$. This shows that if $\lambda _{m} =\pm 1$ then $\mathbf{b}_{m}=\pm \mathbf{a}_{m}$. Therefore if $\lambda _{m} =\pm 1$ then $U_{n}^{FF}\mathbf{a}_{m}=\pm \nu_{1}\mathbf{a}_{m}$.

For cases with $\lambda_{m}\neq \pm 1$, we see from Eq. (\ref{eq:Uab}) that the operator $U_{n}^{FF}$ is a linear operator acting on the linear space $\text{Span}\ (\mathbf{a}_{m}, \mathbf{b}_{m})$. 
In order to obtain the eigenvalues and eigenvectors, we take a vector $\alpha \mathbf{a}_{m} + \beta  \mathbf{b}_{m}\in \text{Span}\ (\mathbf{a}_{m}, \mathbf{b}_{m})$. The eigen equation for $U_{n}^{FF}$ is given by $U_{n}^{FF}(\alpha \mathbf{a}_{m} + \beta  \mathbf{b}_{m}) = \mu (\alpha \mathbf{a}_{m} + \beta  \mathbf{b}_{m})$. From Eq. \eqref{eq:Uab}, this is equivalent to 
\begin{eqnarray*}
\begin{bmatrix}
0 & \nu_{2}\\
\nu_{1} & (\nu_{1}-\nu_{2})\lambda_{m}
\end{bmatrix}
\begin{bmatrix}
\alpha \\
\beta
\end{bmatrix}
=
\mu
\begin{bmatrix}
\alpha \\
\beta
\end{bmatrix}
.
\end{eqnarray*}
Therefore we can obtain two eigenvalues $\mu_{\pm m}$ of $U_{n}^{FF}$ which are related to the eigenvalue $\lambda_{m}$ of $J_{n}^{QW}$ as solutions of the following quadratic equation:
\begin{eqnarray*}
\mu ^{2}-(\nu_{1}-\nu_{2})\lambda_{m}\mu -\nu_{1}\nu_{2}=0.
\end{eqnarray*}
Also we have the corresponding eigenvectors $\nu_{2}\mathbf{a}_{m}+\mu_{\pm m}\mathbf{b}_{m}$ by setting $\alpha = \nu_{2}, \beta = \mu_{\pm m}$.
As a consequence, we obtain the following lemma:

\begin{lem}\label{lem:eigenUformJ}
Let $\lambda_{m}\ (m=0, \ldots ,n-1)$ be the eigenvalues of $J_{n}^{QW}$, then the corresponding eigenvalues $\mu_{\pm m}$ and the eigenvectors $\mathbf{u}_{\pm m}$ of $U_{n}^{FF}$ are the following:

\begin{enumerate}
\item
If $\lambda_{m}=\pm 1$ then $\mu _{m}=\pm \nu_{1}$ and $\mathbf{u}_{m}=\mathbf{a}_{m}$.
\item
If $\lambda_{m}\neq \pm 1$ then $\mu _{\pm m}$ are the solutions of the following quadratic equation:
\begin{eqnarray*}
\mu ^{2}-(\nu_{1}-\nu_{2})\lambda_{m}\mu -\nu_{1}\nu_{2}=0,
\end{eqnarray*}
and $\mathbf{u}_{\pm m}=\nu_{2}\mathbf{a}_{m}+\mu_{\pm m}\mathbf{b}_{m}$. 
\end{enumerate}
\end{lem} 

\begin{rem}
The quadratic equation in Lemma \ref{lem:eigenUformJ} is rearranged to
\begin{align*}
\left\{ i \overline{\nu_{1}}^{1/2} \overline{\nu_{2}}^{1/2} \mu \right\}^{2} + 2\Im (\nu_{1}^{1/2} \overline{\nu_{2}}^{1/2})\lambda_{m}\left\{ i \overline{\nu_{1}}^{1/2} \overline{\nu_{2}}^{1/2} \mu \right\} + 1
&=
0.
\end{align*}
Thus we have
\begin{align*}
i \overline{\nu_{1}}^{1/2} \overline{\nu_{2}}^{1/2} \mu _{\pm m}
&=
- \Im (\nu_{1}^{1/2} \overline{\nu_{2}}^{1/2})\lambda_{m} \pm i \sqrt{1-\left( \Im (\nu_{1}^{1/2} \overline{\nu_{2}}^{1/2})\lambda_{m} \right)^{2}}
\\
\mu _{\pm m}
&=
\left( -\nu_{1}\nu_{2} \right)^{1/2}e^{\pm i \theta _{m}},
\end{align*}
where $\cos \theta_{m} = - \Im (\nu_{1}^{1/2} \overline{\nu_{2}}^{1/2})\lambda_{m}$. 
Therefore if we put $\nu_{j} = e^{i\phi _{j}}$ then the eigenvalues $\mu _{\pm m}$ are given by the following procedure:
\begin{enumerate}
\item
Rescale the eigenvalue $\lambda_{m}$ of $J_{n}^{QW}$ as $- \Im (\nu_{1}^{1/2} \overline{\nu_{2}}^{1/2})\lambda_{m} = - \sin [(\phi_{1} - \phi_{2})/2]\times \lambda _{m}$.
\item
Map the rescaled eigenvalue upward and downward to the unit circle on the complex plane.
\item
Take $[(\phi_{1} + \phi_{2} - \pi )/2]$-rotation of the mapped eigenvalues.
\end{enumerate}
For usual Szegedy walk cases, i.e., $\nu _{1} = 1, \nu_{2} = -1$ case, we have $\phi _{1} = 0, \phi_{2} = \pi$. Thus we can omit 1 and 3 of the procedure because $- \sin [(\phi_{1} - \phi_{2})/2] = 1, [(\phi_{1} + \phi_{2} - \pi )/2] = 0$. 
\end{rem}

\begin{rem}\label{rem:eigenUreminder}
According to Lemma \ref{lem:eigenUformJ}, if all $n$ numbers of eigenvalues of $J_{n}^{QW}$ are not equal to $\pm 1$ then we obtain all $2n$ numbers of eigenvalues of $U_{n}^{FF}$. But if there exist $s$ numbers of the $\lambda_{m}=\pm 1$ eigenvalues of $J_{n}^{QW}$ then we can only obtain $2n-s$ numbers of eigenvalues of $U_{n}^{FF}$. 

In this case, for every $\lambda _{m}=\pm 1$, we construct the following two vectors:
\begin{align*}
\tilde{\mathbf{a}}_{m}
&=
\sum_{i=0}^{n-1}v_{m}(i)\ket{i}\otimes \ket{w_{2,i}},\\
\tilde{\mathbf{b}}_{m}
&=
S_{n}^{FF}\tilde{\mathbf{a}}_{m},
\end{align*}
where $\ket{w_{2,i}}$ is the eigenvector corresponding to the eigenvalue $\nu_{2}$ of $C_{i}$ in Eq. \eqref{specC}. By the definition, we have $|\tilde{\mathbf{a}}_{m}|=|\tilde{\mathbf{b}}_{m}|=1$. Also we obtain the inner product $(\tilde{\mathbf{a}}_{m}, \mathbf{a}_{m})=0$ from orthogonality and $(\tilde{\mathbf{b}}_{m}, \mathbf{a}_{m})=(\tilde{\mathbf{b}}_{m}, \pm\mathbf{b}_{m})=\pm(S_{n}^{FF}\tilde{\mathbf{a}}_{m}, S_{n}^{FF}\mathbf{a}_{m})=0$ from $\lambda _{m}=\pm 1$ and $(S_{n}^{FF})^{2}=I_{n}\otimes I_{2}$. Since $\mathbf{a}_{m}$ belongs to the eigensystem of $\nu _{1}$ of $\mathcal{C}_{n}$, this shows that both $\tilde{\mathbf{a}}_{m}$ and $\tilde{\mathbf{b}}_{m}$ belong to the eigensystem of $\nu _{2}$ of $\mathcal{C}_{n}$. This implies  that 
\begin{eqnarray*}
U_{n}^{FF}
\begin{bmatrix}
\tilde{\mathbf{a}}_{m}\\
\tilde{\mathbf{b}}_{m}
\end{bmatrix}
=
\begin{bmatrix}
0 & \nu_{2}\\
\nu_{2} & 0
\end{bmatrix}
\begin{bmatrix}
\tilde{\mathbf{a}}_{m}\\
\tilde{\mathbf{b}}_{m}
\end{bmatrix}.
\end{eqnarray*}
Therefore $U_{n}^{FF}(\tilde{\mathbf{a}}_{m}\pm \tilde{\mathbf{b}}_{m})=\pm \nu_{2}(\tilde{\mathbf{a}}_{m}\pm \tilde{\mathbf{b}}_{m})$. These are the candidates of eigenvalues and eigenvectors.

On the other hand, the two sets $\mathcal{H}_{n}^{(\pm)}= \mathrm{span}\{\ket{i+1,L}\pm \ket{i,R}:i\in V_{n}\ (\!\!\! \mod n)\}$ are subspaces of whole Hilbert space $\mathcal{H}_{n}$ with $\mathrm{dim}\mathcal{H}_{n}^{(\pm)}=n$, i.e., $\mathcal{H}_{n}=\mathcal{H}_{n}^{(+)}\oplus \mathcal{H}_{n}^{(-)}$. Note that $\mathbf{a}_{m}\pm \mathbf{b}_{m}\in \mathcal{H}_{n}^{(\pm)}$. If $\lambda _{m}=\pm 1$ then $\mathbf{a}_{m}=\pm \mathbf{b}_{m}$. 
This implies that if $\lambda _{m}=\pm 1$ then the dimension of $\mathcal{H}_{n}^{(\mp)} \cap \text{Span}\ (\mathbf{a}_{m}, \mathbf{b}_{m})$ decreases by $1$. 
Therefore if $\lambda _{m}=1$ then we can only choose $U_{n}^{FF}(\tilde{\mathbf{a}}_{m}- \tilde{\mathbf{b}}_{m})=\nu_{2}(\tilde{\mathbf{a}}_{m}- \tilde{\mathbf{b}}_{m})$. In the same way, if $\lambda _{m}=-1$ then we can only choose  $U_{n}^{FF}(\tilde{\mathbf{a}}_{m}+ \tilde{\mathbf{b}}_{m})=-\nu_{2}(\tilde{\mathbf{a}}_{m}+ \tilde{\mathbf{b}}_{m})$. Using these procedure, we have remaining s numbers of eigenvalues and eigenvectors.
\end{rem}

As a consequence of Lemmas \ref{lem:polyJacobi}, \ref{lem:Kreal}, \ref{lem:eigenUformJ} and Remark \ref{rem:eigenUreminder}, we have the following result:

\begin{thm}\label{thm:periodUFFisospectral}
Under the Assumption \ref{ass:coinQW}, let $w_{j}(R)=\sqrt{p_{j}}e^{i\theta _{R}(j)}$ and $w_{j}(L)=\sqrt{q_{j}}e^{i\theta _{L}(j)}$ for $j\in V_{n}$ where $p_{j}=|w_{j}(R)|^{2}$ and $q_{j}=|w_{i}(L)|^{2}$. Also let $\tilde{U}_{n}^{FF}$ which is defined by the coin operator $\tilde{\mathcal{C}}_{n}$ with $\tilde{w}_{j}(R)=\sqrt{p_{j}}e^{i(\theta _{R}(j)+\tilde{\theta }_{R}(j))}$ and $\tilde{w}_{j}(L)=\sqrt{q_{j}}e^{i(\theta _{L}(j)+\tilde{\theta }_{L}(j))}$ for $j\in V_{n}$. If $\sum_{j=1}^{n-1}(\tilde{\theta}_{L}(j)-\tilde{\theta }_{R}(j))=2\pi k\ (k\in \mathbb{Z})$ then $T_{n}(U_{n}^{FF})=T_{n}(\tilde{U}_{n}^{FF})$.
\end{thm}
{\bf Proof of Theorem \ref{thm:periodUFFisospectral}.}

From Eq.\ \eqref{eq:polyJacobi}, if $\sum_{j=1}^{n-1}(\tilde{\theta}_{L}(j)-\tilde{\theta}_{R}(j))=2\pi k\ (k\in \mathbb{Z})$ then $\Re \left(\prod_{j=0}^{n-1}\overline{w_{j}(R)}w_{j}(L)\right)=\Re \left(\prod_{j=0}^{n-1}\overline{\tilde{w}_{j}(R)}\tilde{w}_{j}(L)\right)$. Then from Lemmas \ref{lem:polyJacobi}, \ref{lem:Kreal}, \ref{lem:eigenUformJ} and Remark \ref{rem:eigenUreminder}, we have $\mathrm{Spec}\ U_{n}^{FF} = \mathrm{Spec}\ \tilde{U}_{n}^{FF}$. Therefore we have $T_{n}(U_{n}^{FF})=T_{n}(\tilde{U}_{n}^{FF})$.
\qed

Theorem \ref{thm:periodUFFisospectral} provides a classification of our DTQW from the point of the periodicity. Indeed, $T_{n}(U_{n}^{FF})$ depends only on the sequence $\{p_{j}\}_{0\leq j \leq n-1}$ and a value $\sum_{j=1}^{n-1}(\tilde{\theta}_{L}(j)-\tilde{\theta }_{R}(j))$. Therefore we can identify DTQWs having the same set of these values. The next corollary provides ``Hadamard class'' of periodicity.

\begin{cor}\label{cor:UMSHadamard}
Let $\mathcal{C}_{n}^{\prime}=\sum_{j=0}^{n-1}\ket{j}\bra{j}\otimes C^{\prime}_{j}$ 
with 
$C_{j}^{\prime}
=
\frac{1}{\sqrt{2}}
\begin{bmatrix}
e^{i\tilde{\theta}(j)} & 1
\\
1 & -e^{-i\tilde{\theta}(j)}
\end{bmatrix}
$. 
If $\sum_{j=0}^{n-1}\tilde{\theta}(j)=2\pi k\ (k\in \mathbb{Z})$ then
\begin{align*}
T_{n}(U^{MS}_{n}(\mathcal{C}_{n}^{\prime}))
=
\begin{cases}
2, &(n=2)\\
8, &(n=4)\\
24 &(n=8)\\
\infty &(n\neq 2,4,8).
\end{cases}
\end{align*}
\end{cor}
{\bf Proof of Corollary \ref{cor:UMSHadamard}.}

Let $\mathcal{C}_{n}=\sum_{j=0}^{n-1}\ket{j}\bra{j}\otimes H$ 
with 
$
H
=
\frac{1}{\sqrt{2}}
\begin{bmatrix}
1 & 1
\\
1 & -1
\end{bmatrix}
$, i.e., the Hadamard walk case. The periodicity for this case is as follows \cite{Dukes2014, KonnoShimizuTakei2015}:

\begin{align*}
T_{n}(U^{MS}_{n}(\mathcal{C}_{n}))
=
\begin{cases}
2, &(n=2)\\
8, &(n=4)\\
24 &(n=8)\\
\infty &(n\neq 2,4,8).
\end{cases}
\end{align*}
From Lemma \ref{lem:spec}, we have $T_{n}(U_{n}^{MS}(\mathcal{C}_{n}))=T_{n}(U_{n}^{FF}(\mathcal{C}_{n}\sigma_{x}))$. So we consider 
$
H\sigma_{1} = 
\frac{1}{\sqrt{2}}
\begin{bmatrix}
1 & 1 \\
-1 & 1
\end{bmatrix}
$
case. By direct calculation, we obtain
\begin{align*} 
\frac{1}{\sqrt{2}}
\begin{bmatrix}
1 & 1 \\
-1 & 1
\end{bmatrix}
\begin{bmatrix}
1/\sqrt{2}\\
i/\sqrt{2}
\end{bmatrix}
&=
\frac{1+i}{\sqrt{2}}
\begin{bmatrix}
1/\sqrt{2}\\
i/\sqrt{2}
\end{bmatrix}
=
e^{i\pi /4}
\begin{bmatrix}
1/\sqrt{2}\\
i/\sqrt{2}
\end{bmatrix},
\\
\frac{1}{\sqrt{2}}
\begin{bmatrix}
1 & 1 \\
-1 & 1
\end{bmatrix}
\begin{bmatrix}
1/\sqrt{2}\\
-i/\sqrt{2}
\end{bmatrix}
&=
\frac{1-i}{\sqrt{2}}
\begin{bmatrix}
1/\sqrt{2}\\
-i/\sqrt{2}
\end{bmatrix}
=
e^{-i\pi /4}
\begin{bmatrix}
1/\sqrt{2}\\
-i/\sqrt{2}
\end{bmatrix}.
\end{align*}
Therefore the spectral decomposition of the coin operator $H\sigma_{1}$ is
\begin{align*} 
H\sigma_{1}
&=
(e^{i\pi /4}-e^{-i\pi /4})
\begin{bmatrix}
1/\sqrt{2}\\
i/\sqrt{2}
\end{bmatrix}
\begin{bmatrix}
1/\sqrt{2} & -i/\sqrt{2}
\end{bmatrix}
+
e^{-i\pi /4}I_{2}.
\end{align*}
We consider the coin operator $\tilde{\mathcal{C}}_{n}=\sum_{j=0}^{n-1}\ket{j}\bra{j}\otimes \widetilde{(H\sigma_{1})_{j}}$ with
\begin{align*} 
\widetilde{(H\sigma_{1})_{j}}
&=
(e^{i\pi /4}-e^{-i\pi /4})
\begin{bmatrix}
1e^{i\tilde{\theta}_{L}(j)}/\sqrt{2}\\
ie^{i\tilde{\theta}_{R}(j)}/\sqrt{2}
\end{bmatrix}
\begin{bmatrix}
1e^{-i\tilde{\theta}_{L}(j)}/\sqrt{2} & -ie^{-i\tilde{\theta}_{R}(j)}/\sqrt{2}
\end{bmatrix}
+
e^{-i\pi /4}I_{2}
\\
&=
\frac{i}{\sqrt{2}}
\begin{bmatrix}
1 & -ie^{i(\tilde{\theta}_{L}(j)-\tilde{\theta}_{R}(j))}
\\
ie^{-i(\tilde{\theta}_{L}(j)-\tilde{\theta}_{R}(j))} & 1
\end{bmatrix}
+
\frac{1}{\sqrt{2}}
\begin{bmatrix}
1-i & 0
\\
0 & 1-i
\end{bmatrix}
\\
&=
\frac{1}{\sqrt{2}}
\begin{bmatrix}
1 & e^{i(\tilde{\theta}_{L}(j)-\tilde{\theta}_{R}(j))}
\\
-e^{-i(\tilde{\theta}_{L}(j)-\tilde{\theta}_{R}(j))} & 1
\end{bmatrix}
=
\frac{1}{\sqrt{2}}
\begin{bmatrix}
1 & e^{i\tilde{\theta}(j)}
\\
-e^{-i\tilde{\theta}(j)} & 1
\end{bmatrix}.
\end{align*}
Using Theorem \ref{thm:periodUFFisospectral}, we have $T_{n}(U_{n}^{FF}(\tilde{\mathcal{C}}_{n}))=T_{n}(U_{n}^{FF}(\mathcal{C}_{n}))$ if $\sum_{j=0}^{n-1}\tilde{\theta}(j)=2\pi k\ (k\in \mathbb{Z})$. Noting that $\mathcal{C}_{n}^{\prime}=\tilde{\mathcal{C}}_{n}\sigma _{1}$, we obtain the desired result by Lemma \ref{lem:spec}.
\qed

\begin{rem}
By the same arguments of the proof of Corollary \ref{cor:UMSHadamard}, we obtain the following result:

Let $\mathcal{C}_{n}=\sum_{j=0}^{n-1}\ket{j}\bra{j}\otimes C$ 
with 
$
C
=
\begin{bmatrix}
a & b
\\
c & d
\end{bmatrix}
$ 
and 
$\tilde{\mathcal{C}}_{n}=\sum_{j=0}^{n-1}\ket{j}\bra{j}\otimes \tilde{C}_{j}$ 
with 
$\tilde{C}_{j}
=
\begin{bmatrix}
ae^{i\tilde{\theta}(j)} & b
\\
c & de^{-i\tilde{\theta}(j)}
\end{bmatrix}
$. 
If $\sum_{j=0}^{n-1}\tilde{\theta}(j)=2\pi k\ (k\in \mathbb{Z})$ then $T_{n}(U^{MS}_{n}(\mathcal{C}_{n}))=T_{n}(U^{MS}_{n}(\tilde{\mathcal{C}}_{n}))$. 
\end{rem}

\section{Non-isospectral coin cases}
In this section, we consider several types of DTQWs with non-isospectral coin and the moving shift. In order to define periodic coin operator, we introduce a notation $[C:l, \tilde{C}:m]$ which denotes 
\begin{align*}
C_{i}=
\begin{cases}
C &\text{if $0\leq i\leq l-1\ (\! \! \mod (l+m))$},\\
\tilde{C} &\text{if $l\leq i\leq l+m-1\ (\! \! \mod (l+m))$},
\end{cases}
\end{align*}
in the coin operator $\mathcal{C}_{n}=\sum_{i=0}^{n-1}\ket{i}\bra{i}\otimes C_{i}$. In this section, we consider $[C:1, I_{2}:m]$ model with $n=0\ (\! \! \mod m+1)$ for $2\times 2$ unitary matrix $C$.

At the beginning, we consider $m=n-1$ cases. In this cases, the coin operator is defined by $\mathcal{C}_{n}=\ket{0}\bra{0}\otimes C + \sum_{i=1}^{n-1}\ket{i}\bra{i}\otimes I_{2}$ then the time evolution operator $U_{n}^{MS}=S_{n}^{MS}\mathcal{C}_{n}$ is given by
\begin{align*}
U_{n}^{MS}
&=
\ket{0}\bra{1}\otimes \ket{L}\bra{L}I_{2}+\ket{0}\bra{n-1}\otimes \ket{R}\bra{R}I_{2}\\
&\quad +
\ket{1}\bra{2}\otimes \ket{L}\bra{L}I_{2}+\ket{1}\bra{0}\otimes \ket{R}\bra{R}C\\
&\quad +
\sum_{i=2}^{n-2}(\ket{i}\bra{i+1}\otimes \ket{L}\bra{L}I_{2}+\ket{i}\bra{i-1}\otimes \ket{R}\bra{R}I_{2})\\
&\quad +
\ket{n-1}\bra{0}\otimes \ket{L}\bra{L}C+\ket{n-1}\bra{n-2}\otimes \ket{R}\bra{R}I_{2}\\
&=
\ket{0}\bra{1}\otimes \ket{L}\bra{L}+\ket{0}\bra{n-1}\otimes \ket{R}\bra{R}\\
&\quad +
\ket{1}\bra{2}\otimes \ket{L}\bra{L}+\ket{1}\bra{0}\otimes \ket{R}\bra{R}C\\
&\quad +
\sum_{i=2}^{n-2}(\ket{i}\bra{i+1}\otimes \ket{L}\bra{L}+\ket{i}\bra{i-1}\otimes \ket{R}\bra{R})\\
&\quad +
\ket{n-1}\bra{0}\otimes \ket{L}\bra{L}C+\ket{n-1}\bra{n-2}\otimes \ket{R}\bra{R}.
\end{align*}
Thus we have
\begin{align*}
(\bra{0}\otimes I_{2})\left(U_{n}^{MS}\right)^{kn}(\ket{0}\otimes I_{2})
&=
\ket{L}\bra{L}C^{k}+\ket{R}\bra{R}C^{k}
=
(\ket{L}\bra{L}+\ket{R}\bra{R})C^{k}
=
I_{2}C^{k}
\\
&=
C^{k},
\end{align*}
for $k=1,2,\ldots$. Also for $i\neq 0$, we obtain
\begin{align*}
(\bra{i}\otimes I_{2})\left(U_{n}^{MS}\right)^{kn}(\ket{i}\otimes I_{2})
&=
\left(\ket{L}\bra{L}\right)^{n-i-1}\left(\ket{L}\bra{L}C\right)C^{k-1}\left(\ket{L}\bra{L}\right)^{i}
\\
&\quad +
\left(\ket{R}\bra{R}\right)^{i-1}\left(\ket{R}\bra{R}C\right)C^{k-1}\left(\ket{R}\bra{R}\right)^{n-i}
\\
&=
\ket{L}\bra{L}C^{k}\ket{L}\bra{L}+\ket{R}\bra{R}C^{k}\ket{R}\bra{R}
,
\end{align*}
for $k=1,2,\ldots$. Using this observation, we can reach the following result:
\begin{thm}\label{thm:C1In-1}
For the $[C:1, I_{2}:n-1]$ model for $2\times 2$ unitary matrix $C$, let $\lambda _{1}, \lambda _{2}$ be the pair of eigenvalues of $C$. If $\lambda _{1}=\exp [2\pi i(L_{1}/N_{1})]$ and $\lambda _{2}=\exp [2\pi i(L_{2}/M_{2})]$ where $L_{1}/N_{1}$ and $L_{2}/N_{2}$ are reduced rational numbers, we take $M=l.c.m.(M_{1},M_{2})$ then $T_{n}(U_{n}^{FF})=Mn$, where $l.c.m.(a,b)$ denotes the least common multiple of two integers $a$ and $b$.
\end{thm}

From the above discussion, we can see the vertex which has the coin $I_{2}$ just through the coin state. Therefore we have the following result for general $[C:1, I_{2}:m]$ model with $n=0\ (\! \! \mod m+1)$ for $2\times 2$ unitary matrix $C$:
\begin{cor}\label{cor:C1Im}
For the $[C:1, I_{2}:m]$ model with $n=0\ (\! \! \mod m+1)$ for $2\times 2$ unitary matrix $C$, let $T_{n/(m+1)}^{FF}$ be the period of the time evolution operator with coin operator $\mathcal{C}_{n/(m+1)}=\sum_{i=0}^{n/(m+1)-1}\ket{i}\bra{i}\otimes C$ and flip-flop shift operator $S_{n/(m+1)}^{FF}$. Then we have $T_{n}(U_{n}^{FF})=(m+1)T_{n/(m+1)}^{FF}$.
\end{cor}

\par
\
\par\noindent
{\bf Acknowledgments.} 

This study was partially supported by Yokohama Academic Foundation. C. L. H. was supported in part by the Ministry of Science and Technology (MoST) of the Republic of China under Grants 102-2112-M-032-003-MY3 and 105-2918-I-032-001. Y. I. was supported by the Grant-in-Aid for Young Scientists (B) of Japan Society for the Promotion of Science (Grant No. 16K17652). N. K. was supported by the Grant-in-Aid for Challenging Exploratory Research of Japan Society for the Promotion of Science (Grant No. 15K13443). 

C. L. H. would like to thank T. Deguchi,  E. Uehara, E. Nozawa, C. Matsuyama, and N. Oshima,  for the hospitality extended to him during his visit to the Department of Physics of Ochanomizu University. We also thank the anonymous referees for give us fruitful comments on this paper.

\begin{small}

\end{small}

\end{document}